\documentclass[amsmath,amssymb,showkeys,12pt,tightenlines,nobibnotes,nofootinbib]{revtex4}
\usepackage{amsfonts}
\usepackage{mathrsfs}
\usepackage{graphicx}
\usepackage{amsmath}
\usepackage{amssymb}
\usepackage{amsbsy}
\usepackage{marvosym}
\usepackage{url}
\makeatletter
\newcommand{\linea}{\noindent\rule{1.0\textwidth}{1pt}}%

\usepackage{color}
\makeatother

\begin{document}

\title{Dynamics of asymmetric kinetic Ising systems revisited}

\author{Haiping Huang}
\affiliation{Department of Computational Intelligence and Systems
Science, Tokyo Institute of Technology, Yokohama 226-8502, Japan}
\author{Yoshiyuki Kabashima}
\affiliation{Department of Computational Intelligence and Systems
Science, Tokyo Institute of Technology, Yokohama 226-8502, Japan}
\date{\today}

\begin{abstract}
The dynamics of an asymmetric kinetic Ising model is studied. Two
schemes for improving the existing mean-field description are
proposed. In the first scheme, we derive the formulas for
instantaneous magnetization, equal-time correlation, and
time-delayed correlation, considering the correlation between
different local fields. To derive the time-delayed correlation, we
emphasize that the small correlation assumption adopted in previous
work [M. M\'ezard and J. Sakellariou, J. Stat. Mech., L07001 (2011)]
is in fact not required. To confirm the prediction efficiency of our
method, we perform extensive simulations on single instances with
either temporally constant external driving fields or sinusoidal
external fields. In the second scheme, we develop an improved
mean-field theory for instantaneous magnetization prediction
utilizing the notion of the cavity system in conjunction with a
perturbative expansion approach. Its efficiency is numerically
confirmed by comparison with the existing mean-field theory when
partially asymmetric couplings are present.

\end{abstract}

 \maketitle

\section{Introduction}
The dynamics of asymmetric kinetic Ising systems has been
intensively studied in the statistical physics
community~\cite{Somp-1988,Coolen-1996,Roudi-2011,Mezard-2011,Aurell-2012}.
In equilibrium statistical physics, symmetry is assumed to construct
couplings between spins, which leads to a simple stationary state
described by the Gibbs-Boltzmann distribution~\cite{Saad-2013}.
However, a more realistic case is that couplings between spins are
fully or partially asymmetric; an example has been observed in real
neuronal systems~\cite{Nature-11}, where two neurons do not simply
affect each other in a symmetric way. In this case, the dynamics
still has a stationary state but with a rather complicated form
depending on the details of the model~\cite{Aurell-2012}. Therefore,
the static macroscopic quantities of interest have to be computed in
the long-time limit~\cite{Somp-1988}. Further, studies of such
nonequilibrium systems are relevant to model spatio-temporal
statistics of various biological
systems~\cite{Parisi-1986,Derrida-1987,Marre-2009,Nature-08,Nature-13,Roudi-2013},
in the sense that the time-dependent observables can be predicted at
the current time point according solely to knowledge at the previous
time point. Here, we focus on evaluating time-dependent
magnetizations and equal-time and time-delayed correlations for
different sites in a fully or partially asymmetric kinetic Ising
system with parallel (synchronous) dynamics. In
Ref.~\cite{Mezard-2011}, these observables were already evaluated by
assuming negligible correlations of local fields or correlations
between spins at the same time step. We argue that such a
small-correlation assumption is not necessary to derive a
closed-form equation, and we improve the prediction accuracy of these
time-dependent quantities by incorporating these correlations. Given
the finite system size, we show that the improvement is much more
significant, particularly in the low-temperature region, by
comparing these two mean-field methods.

In general, there exist correlations between couplings; i.e., spins
in the system are partially asymmetrically coupled. As a result,
memory effects become increasingly important, and the theory
developed for fully asymmetric networks~\cite{Mezard-2011} should be
revised by considering the retarded self-interactions induced by the
connection symmetry~\cite{Hatchett-2004}. To this end, we propose an
improved mean-field theory to capture the memory effects and thus
improve the prediction accuracy of time-dependent observables, and we
 support this assertion by numerical simulations on single instances.

The rest of this paper is organized as follows. The asymmetric
kinetic Ising model and the parallel dynamics are introduced in
Sec.~\ref{sec_Model}. Closed-form equations for evaluating
time-dependent quantities such as magnetization, equal-time
correlation, and time-delayed correlation are derived in
Sec.~\ref{sec_Dyn}. Extensive numerical simulations to confirm the
efficiency of our method compared with the method introduced in
Ref.~\cite{Mezard-2011} are performed and discussed. In
Sec.~\ref{sec_IMF}, we develop an improved mean-field theory to
treat the memory effects arising in partially asymmetric connected
networks. Its significance is supported by the numerical simulation
presented in this section. The final section is devoted to a
summary.
\section{Asymmetric kinetic Ising model}
\label{sec_Model}
 The parallel dynamics of a kinetic Ising system is described
by a Markov chain with the transition probability
\begin{equation}\label{TP}
    p(\mathbf{s}(t)|\mathbf{s}(t-1))=\prod_{i=1}^{N}\frac{e^{\beta s_{i}(t)h_{i}(t)}}{2\cosh(\beta h_{i}(t))}
,\end{equation} conditioned to the fact that the $N$-dimensional
Ising spin configuration $\mathbf{s}(t-1)$ at the $(t-1)$th time
step is given. The inverse temperature $\beta$ serves as a measure
of the degree of stochasticity. Parallel dynamics means that the
transition probability for each $s_{i}(t)$ at time $t$ relies only
on the state of its neighbors at time $t-1$. Therefore, we define
the effective field as
$h_{i}(t)\equiv\theta_{i}(t)+\sum_{j\in\partial
i}J_{ij}s_{j}(t-1)$~\cite{Aurell-2012}. $\partial i$ denotes the
neighbors of spin $i$. In the current context, each spin is
connected to other spins; i.e., the cardinality of spin $i$,
$|\partial i|=N-1$. $J_{ij}$ denotes the coupling strength for the
directed edge $(ij)$ from spin $j$ to spin $i$. We assume completely
uncorrelated (fully asymmetric) couplings in the sense that they are
all drawn independently from a Gaussian distribution with zero mean
and variance $1/N$. $\theta_{i}(t)$ refers to the time-dependent
external field and it is chosen to be $\theta_{0}(t)$ or
$-\theta_{0}(t)$ with equal probability for each spin $i$.
Therefore, the parallel dynamics of the asymmetric kinetic Ising
system at a discrete time step is described by the following Glauber
rule for all spins ($i=1,\ldots,N$)~\cite{Derrida-1987,Somp-1988}:
\begin{equation}\label{Glau}
s_{i}(t)=\begin{cases} -1 &\text{with Prob  $1-g(h_{i}(t))$},\\
+1 &\text{with Prob  $g(h_{i}(t))$},
\end{cases}
\end{equation}
where $g(h(t))=(1+e^{-2\beta h(t)})^{-1}$. In the parallel dynamics,
all spins are updated according to Eq.~(\ref{Glau}) simultaneously
at each discrete time step. In the presence of symmetric couplings,
the dynamics will evolve to a simple equilibrium state; however, if
the couplings are asymmetric, the dynamics still has a steady state
but this state is unknown \textit{a priori}. Using features of this
model, we will derive the mean-field equations for instantaneous
macroscopic quantities in the following section and further
demonstrate the difference from the derivation in
Ref.~\cite{Mezard-2011}. We then relax the fully asymmetric
assumption to a partially asymmetric one.

\section{Prediction with correlations between different spins}
\label{sec_Dyn}

Under the transition probability of Eq.~(\ref{TP}), the joint
probability of any spin trajectory
$\mathbf{s}(0),\mathbf{s}(1),\ldots,\mathbf{s}(t)$ is given by
\begin{equation}\label{JP}
    p(\mathbf{s}(0),\mathbf{s}(1),\ldots,\mathbf{s}(t))=p(\mathbf{s}(t)|\mathbf{s}(t-1))\prod_{\sigma=1}^{t-1}p(\mathbf{s}(\sigma)|\mathbf{s}(\sigma-1))P(\mathbf{s}(0)),
\end{equation}
due to the Markovian property. $P(\mathbf{s}(0))$ is the initial
distribution. The instantaneous magnetization is defined as
$m_{i}(t)\equiv\left<s_{i}(t)\right>$, where the average operation $\left<\cdots\right>$ is taken
over the trajectory spin history (i.e., over the path probability Eq.~(\ref{JP}))~\cite{Somp-1988}. We are also
interested in the time evolution of the equal-time correlation and
the time-delayed correlation. They are defined, respectively, as
$C_{ij}(t)\equiv\left<s_{i}(t)s_{j}(t)\right>-m_{i}(t)m_{j}(t)$ and
$D_{ij}(t)\equiv\left<s_{i}(t+1)s_{j}(t)\right>-m_{i}(t+1)m_{j}(t)$.
Using Eq.~(\ref{JP}), we rewrite these macroscopic observables for
the parallel dynamics as~\cite{Kappen-2000}:
\begin{subequations}\label{MQ}
\begin{align}
  m_{i}(t) &= \left<\tanh(\beta h_{i}(t))\right>, \\
  C_{ij}(t) &= \left<\tanh(\beta h_{i}(t))\tanh(\beta h_{j}(t))\right>-m_{i}(t)m_{j}(t), \\
  D_{ij}(t) &= \left<s_{j}(t)\tanh(\beta
  h_{i}(t+1))\right>-\left<\tanh(\beta h_{i}(t+1))\right>m_{j}(t).
  \end{align}
\end{subequations}

In the definition of the effective field, the sum of a large number
of independent random variables can be assumed to follow a Gaussian
distribution from the central limit theorem~\cite{Mezard-2011},
because of the fully asymmetric and connected property of the model.
As a result, the distribution of the local field
$\tilde{h}_{i}(t-1)\equiv\sum_{j\in\partial i}J_{ij}s_{j}(t-1)$ is
characterized by its mean and variance. The mean is given by
$a_{i}(t-1)=\sum_{j\in\partial i}J_{ij}m_{j}(t-1)$ and correlation
between two local fields reads
\begin{equation}\label{var}
\left<\tilde{h}_{i}(t)\tilde{h}_{j}(t)\right>-\left<\tilde{h}_{i}(t)\right>\left<\tilde{h}_{j}(t)\right>=\sum_{k,l}J_{il}J_{jk}C_{kl}(t)
=[\mathbf{J}\mathbf{C}\mathbf{J}^{T}]_{ij}\equiv\Delta_{ij}(t).
\end{equation}
With this Gaussian approximation, the trajectory history average in
Eq.~(\ref{MQ}) can be transformed into an integral over the Gaussian
distribution, resulting in the following magnetization and
equal-time correlation:
\begin{subequations}\label{MQG}
\begin{align}
  m_{i}(t) &= \int Dz\tanh\beta(\theta_{i}(t)+a_{i}(t-1)+\sqrt{\Delta_{ii}}z),\label{Mqa} \\
  \begin{split}
  C_{ij}(t) &= \int Dz\int
  Dx\tanh\beta(\theta_{i}(t)+a_{i}(t-1)+\sqrt{\Delta_{ii}-\Delta_{ij}}x+\sqrt{\Delta_{ij}}z)\\
  &\times\int
  Dy\tanh\beta(\theta_{j}(t)+a_{j}(t-1)+\sqrt{\Delta_{jj}-\Delta_{ij}}y+\sqrt{\Delta_{ij}}z)-m_{i}(t)m_{j}(t),\label{Mqb}
  \end{split}
  \end{align}
\end{subequations}
where $Dz\equiv e^{-z^{2}/2}dz/\sqrt{2\pi}$ and we omit the time
index $(t-1)$ for all field covariances. In Eq.~(\ref{Mqb}), if
$\Delta_{ij}<0$, it should be replaced by $-\Delta_{ij}$ and only
$z$ in the first $\tanh(\cdot)$ is replaced by $-z$ to retain
correct covariance between local fields. Note that $\Delta_{ii}$ in
the above equations was treated as $\sum_{j\in\partial
i}J_{ij}^{2}(1-m_{j}^{2}(t-1))$ in Ref.~\cite{Mezard-2011}. We call
this simplified method MF (mean field). Here, we keep the entire
knowledge of the equal-time correlation and expect to improve the
prediction especially in the low-temperature region.
Correspondingly, our method is called MFcorre (mean field with
correlations). A similar idea was also proposed in a recent
interesting work~\cite{Moudi-2013}, where the covariance of local
fields could be recursively determined. Here we use directly the
entire knowledge of the equal-time correlation to compute the field
covariance. Furthermore, the time-delayed correlation derived in
Ref.~\cite{Mezard-2011} can be recovered without any
small-correlation (of local fields) assumption. This is shown by the following
derivation:
\begin{equation}\label{Delay}
    \begin{split}
    \sum_{k}J_{jk}D_{ik}&=\left<\tilde{h}_{j}(t)\tanh\beta(\theta_{i}(t+1)+\tilde{h}_{i}(t))\right>-a_{j}(t)\left<\tanh\beta(\theta_{i}(t+1)+\tilde{h}_{i}(t))\right>\\
    &=\left<\delta a_{j}(t)\tanh\beta(\theta_{i}(t+1)+a_{i}(t)+\delta
    a_{i}(t))\right>\\
    &=\int Dz\int Dx\int
    Dy(\sqrt{\Delta_{jj}-\Delta_{ij}}y+\sqrt{\Delta_{ij}}z)\\
    &\times\tanh\beta(\theta_{i}(t+1)+a_{i}(t)+\sqrt{\Delta_{ii}-\Delta_{ij}}x+\sqrt{\Delta_{ij}}z)\\
    &=\beta\Delta_{ij}\int Dx\int
    Dz(1-\tanh^{2}\beta(\theta_{i}(t+1)+a_{i}(t)+\sqrt{\Delta_{ii}-\Delta_{ij}}x+\sqrt{\Delta_{ij}}z))\\
    &=\beta\Delta_{ij}\int
    D\hat{z}(1-\tanh^{2}\beta(\theta_{i}(t+1)+a_{i}(t)+\sqrt{\Delta_{ii}}\hat{z})),
    \end{split}
\end{equation}
where $\delta a_{i}(t)\equiv\tilde{h}_{i}(t)-a_{i}(t)$. Note that
$\left<\delta a_{i}(t)\delta a_{j}(t)\right>=\Delta_{ij}(t)$, and
all field covariances in Eq.~(\ref{Delay}) have time index $(t)$.
From the third to fourth equality, we used the identity $\int Dz
zF(z)=\int Dz F'(z)$. When arriving at the final equality, we made
the transformation
$\hat{z}=\sqrt{\Delta_{ii}-\Delta_{ij}}x+\sqrt{\Delta_{ij}}z$ (where
$\hat{z}$ follows a Gaussian distribution with zero mean and
variance $\Delta_{ii}$). Finally, we recover the formula for
evaluating the time-delayed correlation as
$\mathbf{D}(t)=\mathbf{A}(t)\mathbf{J}\mathbf{C}(t),$ which has been
derived in Ref.~\cite{Mezard-2011} by discarding terms of order
$\mathcal {O}(\Delta_{ij}^{2})$ when calculating the average.
$\mathbf{A}(t)$ is a diagonal matrix with diagonal terms
$\mathbf{A}_{ii}=\beta\int
D\hat{z}(1-\tanh^{2}\beta(\theta_{i}(t+1)+a_{i}(t)+\sqrt{\Delta_{ii}}\hat{z}))$.
We remark here that the only assumption we used is the Gaussian
approximation, which is guaranteed by the fully asymmetric and
connected properties of the kinetic Ising model under consideration.
In this sense, the equations derived above for time-dependent
macroscopic observables are exact even in the low-temperature
region.

The fully asymmetry constraint can be relaxed to a partially
asymmetric one by introducing correlations for couplings. In this
case, the central limit theorem becomes invalid due to the presence
of correlated couplings. Thus, the above derived equations can only
be used as a crude approximation. The effects of coupling asymmetry
were studied in Ref.~\cite{Mezard-2012}. We applied the same
construction as that in Refs.~\cite{Somp-1988,Mezard-2012}, i.e.,
$J_{ij}=J_{ij}^{s}+kJ_{ij}^{as}$, where $k\geq0$ specifies the
asymmetry degree of couplings. $J_{ij}^{s}=J_{ji}^{s}$ and
$J_{ij}^{as}=-J_{ji}^{as}$, where they follow a Gaussian
distribution with zero mean and variance $\frac{J^{2}}{N(1+k^{2})}$.
We choose $J=1$ here. According to the construction, we have
$\left<J_{ij}J_{ji}\right>=\frac{1-k^{2}}{1+k^{2}}\frac{J^{2}}{N}$,
such that $k=0$ corresponds to a fully symmetric network, while a
fully asymmetric network has $k=1$. To evaluate the instantaneous
equal-time correlation, Eq.~(\ref{Mqb}) may not be used directly,
because $\Delta_{ii}-\Delta_{ij}$ or $\Delta_{jj}-\Delta_{ij}$ may
become negative, which never happens when $k=1$. Instead, for
$k\neq1$, one can use the following approximation
\begin{equation}\label{HTexp}
\begin{split}
    \int
  Dx\tanh\beta(\theta_{i}(t)+a_{i}(t-1)+\sqrt{\Delta_{ii}-\Delta_{ij}}x+\sqrt{\Delta_{ij}}z)\\
  \simeq\tanh\beta\left[\theta_{i}(t)+a_{i}(t-1)+\sqrt{\Delta_{ij}}z
  -\beta m_{i}(t)(\Delta_{ii}-\Delta_{ij}+\Delta_{ij}z^{2})\right]
  \end{split}
\end{equation}
based on a small-coupling expansion~\cite{Aurell-2012}. A similar
approximation can be applied to the $y$-term in Eq.~(\ref{Mqb}). Another possible way is to re-write the Gaussian random number
dependent terms as
\begin{equation}\label{Cres}
\begin{split}
   C_{ij}(t)&= \int Dz\int
  Dx\tanh\beta(\theta_{i}(t)+a_{i}(t-1)+v_{1}x+vz)\\
  &\times\tanh\beta(\theta_{j}(t)+a_{j}(t-1)+v_{2}x+vz)-m_{i}(t)m_{j}(t),
  \end{split}
\end{equation}
where
$v_{1}=\frac{\Delta_{ii}-\Delta_{ij}}{\sqrt{\Delta_{ii}+\Delta_{jj}-2\Delta_{ij}}}$,
$v_{2}=-\frac{\Delta_{jj}-\Delta_{ij}}{\sqrt{\Delta_{ii}+\Delta_{jj}-2\Delta_{ij}}}$
and
$v=\sqrt{\frac{\Delta_{ii}\Delta_{jj}-\Delta_{ij}^{2}}{\Delta_{ii}+\Delta_{jj}-2\Delta_{ij}}}$.
In our simulations, this expression caused no problems, keeping both
$\Delta_{ii}+\Delta_{jj}-2\Delta_{ij}$ and
$\Delta_{ii}\Delta_{jj}-\Delta_{ij}^{2}$ positive. As far as we
investigated, eqs.~(\ref{HTexp}) and~(\ref{Cres}) yielded similar
prediction errors at all temperatures. When cross correlation starts
to have significant contributions to the field covariance (this does
happen in the low temperature regime), prediction of MF, which
incorporates only the auto-correlation, is supposed to have quite
large errors whichever formula (eq.~(\ref{HTexp}) or~(\ref{Cres}))
is employed.

In the numerical simulation, we predict the instantaneous
macroscopic quantities at the current time point based on the
knowledge (data) of the previous time point, using the equations
derived in this section. To test the prediction performance, we
compare the prediction result with that obtained by Monte Carlo
simulations (denoted by ${\rm exp}$), and the performance is
evaluated using the
 root-mean-squared errors
\begin{subequations}\label{rms}
\begin{align}
  \Delta_{m} &= \sqrt{\frac{1}{N}\sum_{i=1}^{N}(m_{i}(t)-m^{{\rm exp}}_{i}(t))^{2}},\label{Merror} \\
  \Delta_{C} &= \sqrt{\frac{1}{N^{2}}\sum_{i,j}^{N}(C_{ij}(t)-C^{{\rm exp}}_{ij}(t))^{2}}, \\
  \Delta_{D} &= \sqrt{\frac{1}{N^{2}}\sum_{i,j}^{N}(D_{ij}(t)-D^{{\rm exp}}_{ij}(t))^{2}}.
  \end{align}
\end{subequations}

\begin{figure}
\centering
    \includegraphics[bb=18 18 292 216,width=0.8\textwidth]{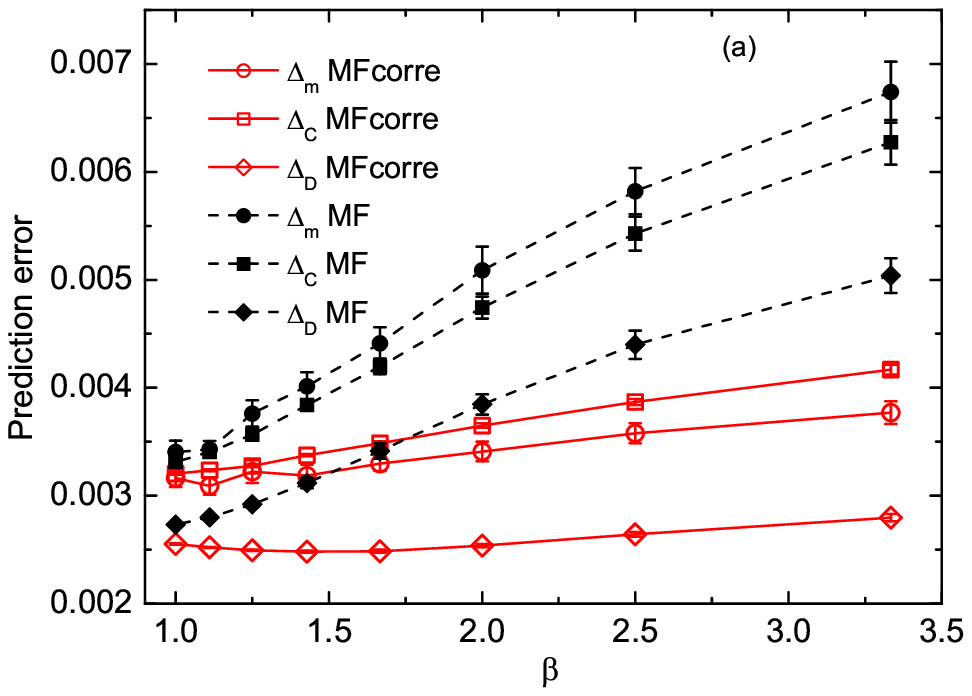}\vskip .1cm
    \includegraphics[bb=17 10 293 220,width=0.8\textwidth]{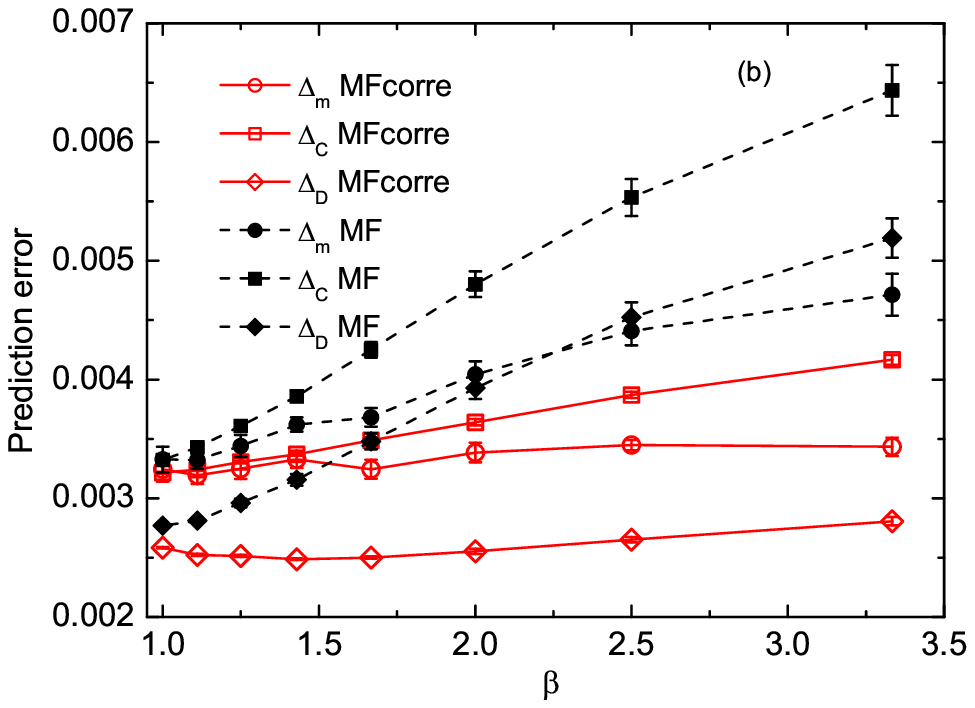}\vskip .1cm
  \caption{
  (Color online) Comparison of prediction performance of MFcorre and MF on fully asymmetric kinetic Ising systems of system size $N=100$.
  Each data point is the average over ten random realizations.
A total number of $10^{5}$ spin trajectories are collected up to
$31$ time steps and these trajectory data are used either to compute
inputs for prediction equations or to compute the experimental values
for comparison. We predict the instantaneous magnetization,
equal-time correlation, and time-delayed correlation at $t=31$ based
on the data at $t=30$. (a) Constant external fields with
$\theta_{0}(t)=0.1$. (b) Sinusoidal external fields with
$\theta_{0}(t)=0.1\sin(2\pi t/t_{0})$. The
 period ($t_{0}$) is chosen to be $10$ time steps.
  }\label{Simu}
\end{figure}

\begin{figure}
\centering
    \includegraphics[bb=15 17 294 218,width=0.8\textwidth]{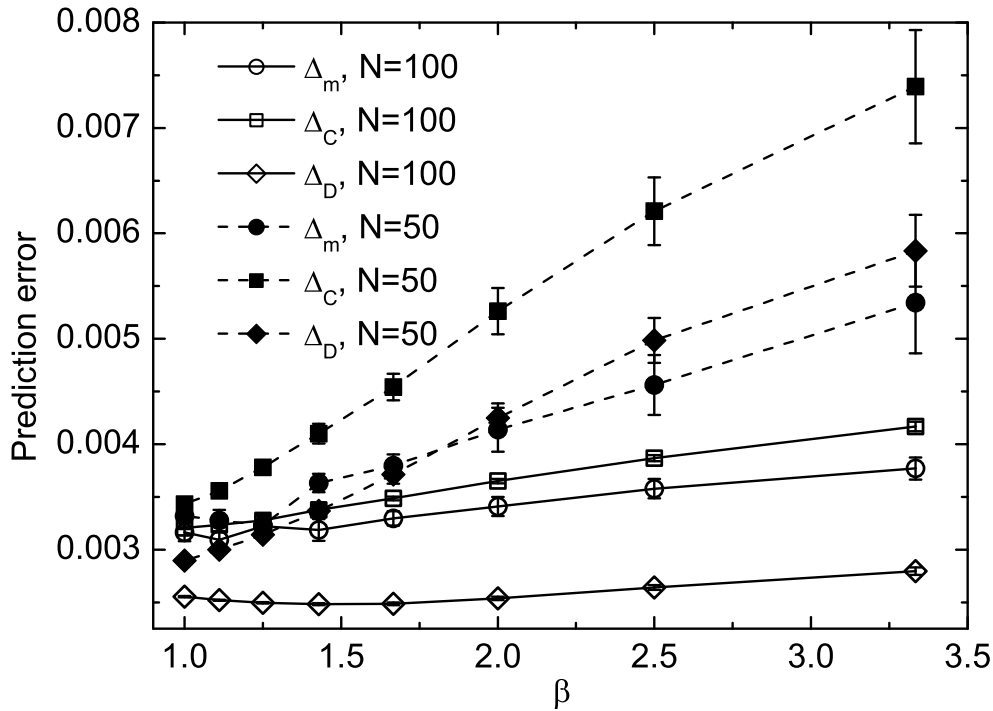}
  \caption{
  (Color online) Finite size dependence of the prediction performance of MFcorre. The same as figure~\ref{Simu}, but for constant external fields and different system
  sizes.
  }\label{MFsize}
\end{figure}

\begin{figure}
\centering
    \includegraphics[bb=16 21 292 220,width=0.6\textwidth]{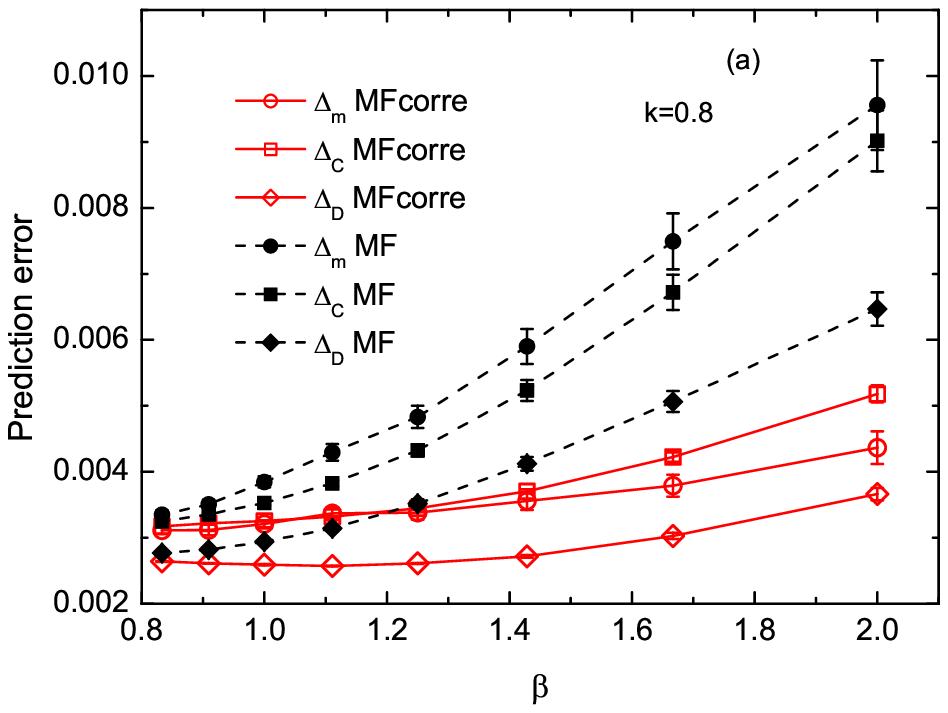}\vskip .1cm
    \includegraphics[bb=18 11 288 216,width=0.6\textwidth]{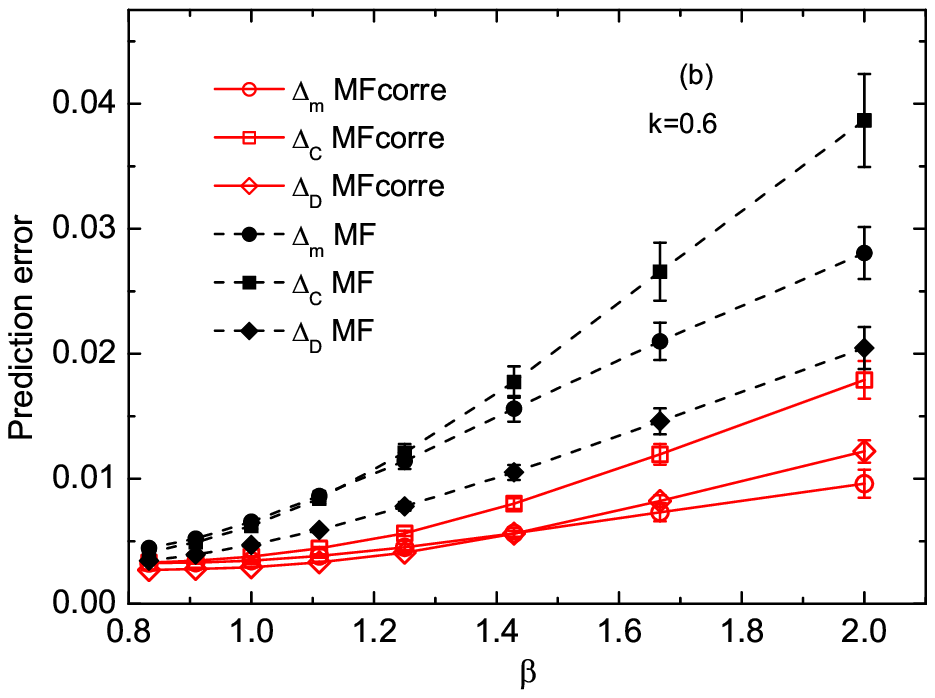}\vskip .1cm
    \includegraphics[bb=16 19 287 217,width=0.6\textwidth]{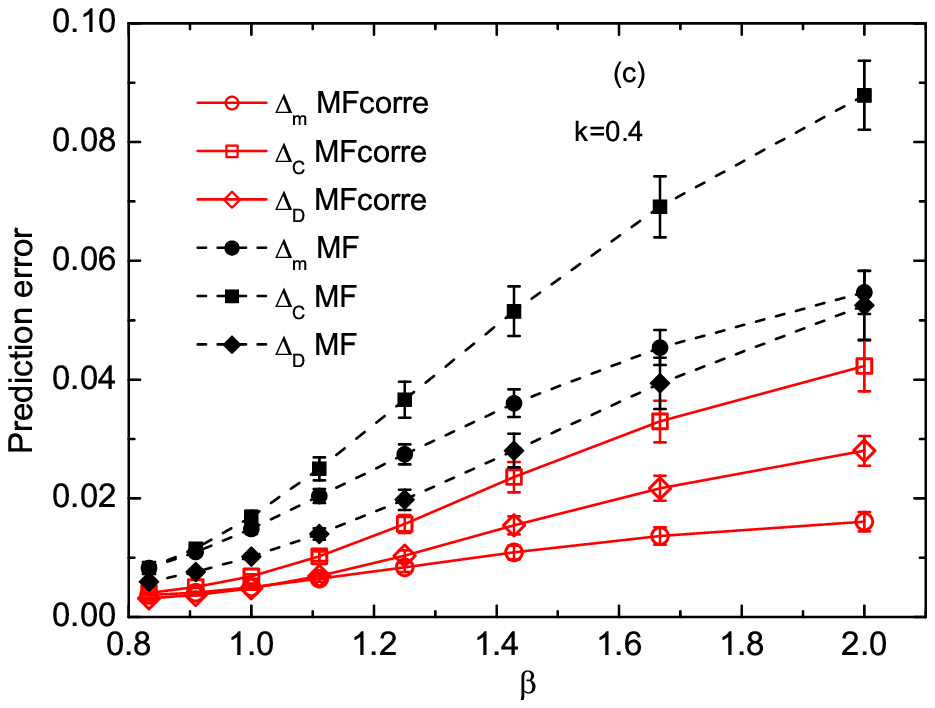}\vskip .1cm
  \caption{
  (Color online) The same as figure~\ref{Simu}, but for constant external fields and different asymmetry degrees $k$. (a) $k=0.8$. (b)
  $k=0.6$. (c) $k=0.4$.
  }\label{Simu02}
\end{figure}

We simulated asymmetric kinetic Ising systems of system size $N=100$
by following the Glauber rule defined in Eq.~(\ref{Glau}). The
initial distribution is chosen such that each spin is randomly
independently assigned $+1$ or $-1$. The dynamics is run up to $31$
time steps with $10^{5}$ spin trajectories. Therefore, a total of
$10^{5}$ instantaneous spin configurations at $t=30$ and $t=31$ are
collected, respectively, where the trajectory data at $t=30$ are
used to compute inputs (magnetizations and equal-time correlations
at time $t=30$) for predicting the time-dependent macroscopic
quantities at $t=31$, and the trajectory data at $t=31$ are used to
compute the experimental values for comparison. The prediction error
reported in Fig.~\ref{Simu} is averaged over ten different random
realizations (coupling constructions and external fields).
Temporally constant and time-varying external fields are applied. As
observed in Fig.~\ref{Simu}, MFcorre improves upon MF, especially in
the low-temperature region, where thermal fluctuations are not
strong and pairwise correlation dominates the dynamic behavior.
Moreover, the prediction error of MFcorre increases slowly with the
inverse temperature, whereas the prediction error of MF increases
rapidly as the temperature decreases. The result implies that, to
achieve perfect accuracy of prediction, incorporating the correlation
is necessary especially for low temperatures and finite system size
of order $\mathcal {O}(10^{2})$, for which the computational cost is
tolerant. We also show the prediction performance of MFcorre on networks
of different sizes in Fig.~\ref{MFsize}, which illustrates that the
prediction error increases for smaller size network, especially at
low temperatures.

The effects of coupling asymmetry on the prediction performance of
the two mean-field methods are summarized in Fig.~\ref{Simu02}. The
performance for both methods degrades as $k$ decreases; however,
MFcorre still outperforms MF in the low-temperature region,
suggesting that considering correlations can compensate for the
prediction error induced by the partial asymmetry, even if the
central limit theorem is applied to derive the prediction equations.

\section{Improved mean-field theory}
\label{sec_IMF} In the previous section, we improved the dynamical
prediction by incorporating the correlation between two different
sites. However, it increases necessary computational time of
prediction particularly for large systems. Furthermore, in
Eq.~(\ref{Mqa}) memory effects induced by connection symmetry have
not been considered, which leads to a high prediction error already
reported in Ref.~\cite{Mezard-2012}. In the case of sparsely coupled
systems, one can keep track of the directed influence from neighbors
by effectively modifying the external field of each spin along the
dynamics~\cite{Neri-2009,Aurell-2011}. However, the direct employment of the
scheme to fully coupled systems requires significant computational
cost and is practically infeasible with current standard
computational resources.

To overcome such a situation, we develop an improved mean-field
theory (IMF) applicable to fully coupled networks with connection
correlations but retain the low complexity of the prediction
algorithm. Here, we treat the memory effect explicitly by
introducing an additional field that describes a backaction from the
states at earlier time steps. In the following derivation, we assume
that couplings are drawn with correlations specified by the
asymmetry degree $k$ introduced in Sec.~\ref{sec_Dyn}. Combining
Eqs.~(\ref{TP}) and (\ref{JP}), we have the joint probability of
spin trajectory given by
\begin{equation}\label{JPIMF}
    p(\mathbf{s}(0),\mathbf{s}(1),\ldots,\mathbf{s}(t))=P(\mathbf{s}(0))\prod_{\sigma=1}^{t}\prod_{l}\frac{e^{\beta s_{l}(\sigma)h_{l}(\sigma)}}{2\cosh(\beta
    h_{l}(\sigma))}.
\end{equation}
We then separate the spin-$i$-related term $J_{li}s_{i}(\sigma-1)$
in the local field $h_{l}(\sigma)$ to consider its directed
influence over its neighbors, and we make the following expansion:
\begin{equation}\label{exp}
    \exp\left[-\ln2\cosh(\beta h_{l}^{(i)}(\sigma)+\beta
    J_{li}s_{i}(\sigma-1))\right]\simeq\frac{\exp\left[-\beta\tanh(\beta h_{l}^{(i)}(\sigma))
    J_{li}s_{i}(\sigma-1)\right]}{2\cosh(\beta
h_{l}^{(i)}(\sigma))},
\end{equation}
where we are allowed to truncate the expansion with respect to
$\beta J_{li} s_i(\sigma-1)$ up to the first order due to its
weakness and the statistical independence among different indices of
$J_{li}$. $h_{l}^{(i)}(t)$ defines the cavity local field as
$h_{l}^{(i)}(t)\equiv\theta_{l}(t)+\sum_{j\in\partial l\backslash
i}J_{lj}s_{j}(t-1)$ where $\backslash i$ indicates that node $i$ is
excluded. Note that this expansion puts a less stringent constraint
on the strength of couplings than that used to derive the dynamical
TAP equation in Refs.~\cite{Roudi-2011,Mezard-2011,Aurell-2012}.

Applying Eq.~(\ref{exp}) in Eq.~(\ref{JPIMF}) completely decouples
the contribution of the trajectory of spin $i$ from the joint
distribution of the cavity system as
\begin{equation}\label{JPIMF-2}
    p(\mathbf{s}(0),\mathbf{s}(1),\ldots,\mathbf{s}(t))\simeq P(\mathbf{s}(0))\prod_{\sigma=1}^{t}\prod_{l\neq i}
    \frac{e^{\beta s_{l}(\sigma)h_{l}^{(i)}(\sigma)}}{2\cosh(\beta h_{l}^{(i)}(\sigma))}\times\prod_{\sigma=1}^{t}
    \frac{e^{\beta s_{i}(\sigma)h_{i}(\sigma)}}{2\cosh(\beta h_{i}(\sigma))}\times\prod_{\sigma=1}^{t}e^{\beta s_{i}(\sigma-1)\phi_{i}(\sigma-1)},
\end{equation}
where the additional field $\phi_{i}(t-1)\equiv\sum_{l\neq
i}J_{li}(s_{l}(t)-\tanh\beta h_{l}^{(i)}(t))$ was introduced. Let us
denote $\mathbf{s}^{(i)}(\sigma)$ as the set of spins at time
$\sigma$ except for $s_i(\sigma)$. Due to the nature of the current
model, both $J_{li}$ and $J_{il}$ are statistically independent of
the cavity distribution
$p^{(i)}(\mathbf{s}^{(i)}(0),\mathbf{s}^{(i)}(1),\ldots,\mathbf{s}^{(i)}(t))
=P^{(i)}(\mathbf{s}^{(i)}(0)) \prod_{\sigma=1}^t \prod_{l\ne i}
\frac{e^{\beta s_{l}(\sigma)h_{l}^{(i)}(\sigma)}}{2\cosh(\beta
h_{l}^{(i)}(\sigma))}$, where $P^{(i)}(\mathbf{s}^{(i)}(0))$ denotes
the joint distribution of the cavity system at $t=0$. Hereafter, we
assume that the initial state is described by a factorized
distribution $P(\mathbf{s}(0))=\prod_{i=1}^{N} P_{i}(s_{i}(0))$, so
that the joint cavity distribution is given as
$P^{(i)}(\mathbf{s}^{(i)}(0))=\prod_{j\neq i} P_{j}(s_{j}(0))$.
This, in conjunction with the central limit theorem, enables us to
handle the field distribution
\begin{eqnarray}
&&\Psi(\boldsymbol{\phi}_{i},\mathbf{h}_{i}) =
\sum_{\mathbf{s}^{(i)}(0), \mathbf{s}^{(i)}(1),\ldots,
\mathbf{s}^{(i)}(t)}
p^{(i)}(\mathbf{s}^{(i)}(0),\mathbf{s}^{(i)}(1),\ldots,\mathbf{s}^{(i)}(t))
\cr && \hspace*{2cm}\times \prod_{\sigma=1}^t \delta \left
(h_i(\sigma)-\theta_i(\sigma)-\sum_{j \ne  i} J_{ij}s_j(\sigma-1)
\right ) \cr && \hspace*{2cm}\times \prod_{\sigma=0}^{t-1}
\delta\left (\phi_{i}(\sigma)-\sum_{l\neq
i}J_{li}(s_{l}(\sigma+1)-\tanh (\beta h_{l}^{(i)}(\sigma+1))) \right
), \label{field_dist}
\end{eqnarray}
as of the Gaussian form with the property that the original local
field $\mathbf{h}_{i} =(h_i(1),h_i(2),\ldots,h_i(t))$ and the
additional backaction field
$\boldsymbol{\phi}_{i}=(\phi_i(0),\phi_i(1),\ldots,\phi_i(t-1))$
have correlations.  This, in conjunction with the last product in
Eq. (\ref{JPIMF-2}), incorporates the memory effect induced by
retarded self-interaction via the cavity system to the $i$-th spin,
which can be understood by the fact that the dynamics of spin $i$ at
earlier time steps will affect the current state of its neighbors.
Equations (\ref{JPIMF-2}) and (\ref{field_dist}) mean that the
marginal distribution of the
 trajectory of spin $i$ can be written as
\begin{eqnarray}\label{IMF-marg}
&& p(s_{i}(0),s_{i}(1),\ldots,s_{i}(t)) =\sum_{\mathbf{s}^{(i)}(0),
\mathbf{s}^{(i)}(1),\ldots,
\mathbf{s}^{(i)}(t)}p(\mathbf{s}(0),\mathbf{s}(1),\ldots,\mathbf{s}(t))\cr
&& = \mathcal {N}^{-1}\int
\mathrm{d}\boldsymbol{\phi}_{i}\mathrm{d}\mathbf{h}_{i}\Psi(\boldsymbol{\phi}_{i},\mathbf{h}_{i})
P_{i}(s_{i}(0) )\cr && \times\exp\left[\sum_{\sigma=1}^{t}\beta
s_{i}(\sigma)h_{i}(\sigma)
    +\sum_{\sigma=1}^{t}\beta s_{i}(\sigma-1)\phi_{i}(\sigma-1)-\sum_{\sigma=1}^{t}\ln2\cosh\beta
    h_{i}(\sigma)\right],
\end{eqnarray}
where $\mathcal {N}$ is a normalization constant. Writing
Eq.~(\ref{IMF-marg}) has the advantage that we can directly take
into account the contribution of backaction field
$\boldsymbol{\phi}_{i}$ in deriving time-dependent quantities of
interest.

To consider the memory effect, we should have data at least up to
two time steps earlier (e.g., $\mathbf{m}(t-2)$). This was also
observed in the dynamical inference in a diluted partially
asymmetric Ising system for which the dynamic cavity
method~\cite{Aurell-2011,Aurell-2012} is computationally feasible.
For the following derivation, we define $\eta_{i}(t)\equiv
h_{i}(t)-\left<h_{i}(t)\right>^{(i)}$, where the superscript $(i)$
means the average is taken without the backaction of spin $i$.
$\left<h_{i}(t)\right>^{(i)}$ can be calculated indirectly as we
shall show. For brevity, the time index for the field is neglected
as $\eta_{i}\equiv\eta_i(t)$ and $\phi_{i}\equiv\phi_{i}(t-2)$. As a
first approximation, we here consider the field correlations only
for this time difference. This is reasonable because both fields are
determined by the state of the cavity network at the same time slice
$t-1$. Improving the approximation level by considering more time
steps is also possible, although the necessary treatment would
become more complicated technically.

The approximation is constructed by handling the state of the
$t-2$-th step as if it were the initial state in Eq.
(\ref{IMF-marg}). This allows us to carry out the summation over
$s_{i}(t-1)$ and integration over $\phi_{i}(t-1)$ independently of
the other relevant variables, which yields an expression
\begin{eqnarray}\label{IMF-marg2}
&&p(s_i(t-2),s_i(t)) =\sum_{s_i(t-1), \mathbf{s}^{(i)}(t-2),
\mathbf{s}^{(i)}(t-1), \mathbf{s}^{(i)}(t)}
p(\mathbf{s}(t-2),\mathbf{s}(t-1),\mathbf{s}(t))\cr && \simeq
{\cal N}^{-1} \int \mathrm{d}\eta_i  \mathrm{d} \phi_i
\Psi(\phi_i,\eta_i) p(s_i(t-2)) \cr && \times \exp \left [  \beta
(\eta_i+ \left \langle h_i \right \rangle^{(i)}) s_i(t) + \beta
\phi_i s_i(t-2) - \ln2\cosh\beta
    (\eta_i+ \left \langle h_i \right \rangle^{(i)}) \right].
\end{eqnarray}
Let us denote $p(s_i(\sigma))=\frac{1+m_i(\sigma) s_i(\sigma)}{2}$.
In addition,  we rewrite the joint distribution of the fields as
$\Psi(\phi_i,\eta_i)=\Psi(\eta_i|\phi_i)\Psi(\phi_i)$, where
\begin{eqnarray}\label{psi_eta_phi}
\Psi(\eta_{i}|\phi_{i})=\frac{1}{\sqrt{2\pi
V_{\eta_{i}|\phi_{i}}}}\exp\left[-\frac{1}{2V_{\eta_{i}|\phi_{i}}}\left
(\eta_{i}-\frac{V_{\eta_{i}\phi_{i}}}{V_{\phi_{i}\phi_{i}}}
\phi_{i}\right )^{2}\right],
\end{eqnarray}
\begin{eqnarray}\label{psi_phi}
\Psi(\phi_{i})=\frac{1}{\sqrt{2 \pi V_{\phi_{i}\phi_{i}}}} \exp
\left [ -\frac{1}{2V_{\phi_{i}\phi_{i}}} \phi_i^2 \right ].
\end{eqnarray}
$V_{\phi_i \phi_i}$, $V_{\eta_i \phi_i}$ and $V_{\eta_i|\phi_i}$
parameterize the variance of $\phi_i$, the covariance between
$\phi_i$ and $\eta_i$, and the conditional variance of $\eta_i$
given $\phi_i$, respectively. By using these, the variance of
$\eta_i$, $V_{\eta_i\eta_i}$, is given as $V_{\eta_i \eta_i}=
V_{\eta_i |\phi_i}+\frac{V_{\eta_i \phi_i}^2}{V_{\phi_i\phi_i}}$.
Equations (\ref{IMF-marg2})--(\ref{psi_phi}) provide the expression
of instantaneous magnetization as
\begin{equation}\label{IMF-mag}
\begin{split}
    m_{i}(t)&=\sum_{s_{i}(t),s_{i}(t-2)}\int
    \mathrm{d}\phi_{i}\mathrm{d}\eta_{i}\frac{e^{\beta(\eta_{i}+\left<h_{i}\right>^{(i)})s_{i}(t)}}{2\cosh(\beta(\eta_{i}+\left<h_{i}\right>))}s_{i}(t)\Psi(\eta_{i}|\phi_{i})
    \Psi(\phi_{i}|s_{i}(t-2))p(s_{i}(t-2))\\
    &=\sum_{s_{i}(t-2)}\int
    Dz\tanh\beta\left[\left<h_{i}\right>^{(i)}+\beta V_{\eta_{i}\phi_{i}}s_{i}(t-2)+\sqrt{V_{\eta_{i}\eta_{i}}}z\right]\frac{1+m_{i}(t-2)s_{i}(t-2)}{2}\\
    &=\sum_{s_{i}(t-2)}\frac{1+m_{i}(t-2)s_{i}(t-2)}{2}\int
    Dz\tanh\beta\Xi(z,s_{i}(t-2)),
    \end{split}
\end{equation}
where we defined the conditional distribution of $\phi_i$ given
$s_i(t-2)$ as $\Psi(\phi_{i}|s_i(t-2)) \propto \Psi(\phi_{i})
\exp\left (\beta \phi_i s_i(t-2) \right )$, and
$\Xi(z,s_{i}(t-2))\equiv\theta_{i}(t)+\sum_{j\in\partial
i}J_{ij}m_{j}(t-1)-\beta
V_{\eta_{i}\phi_{i}}(m_{i}(t-2)-s_{i}(t-2))+\sqrt{V_{\eta_{i}\eta_{i}}}z$.
Note that to get the final expression, an equation to evaluate the
cavity average from the full averages
\begin{eqnarray}\label{full-cavity}
\left<h_{i}\right>^{(i)}=\theta_{i}(t)+\sum_{j\in\partial
i}J_{ij}m_{j}(t-1)-\beta V_{\eta_{i}\phi_{i}}m_{i}(t-2)
\end{eqnarray}
was employed. This equation is derived by combining two relations
$\left \langle h_i \right \rangle = \theta_i(t)+\sum_{j\in\partial
i} J_{ij} m_j(t-1)=\left \langle h_i \right \rangle^{(i)}+\left
\langle \eta_i \right \rangle$ and $\left \langle \eta_{i}\right
\rangle= \sum_{s_i(t-2)}\int \mathrm{d}\phi_i \mathrm{d} \eta_i
\eta_i \Psi(\eta_i|\phi_i) \Psi(\phi_i|s_i(t-2))p(s_i(t-2))=\beta
V_{\eta_{i}\phi_{i}}m_{i}(t-2)$. The last term of
Eq.~(\ref{full-cavity}) indicates subtraction of the retarded
self-interaction effect.

Equations (\ref{IMF-mag}) and (\ref{full-cavity}) indicate that
assessing the second moments of the cavity fields $V_{\eta_i
\eta_i}$ and $V_{\eta_i\phi_i}$ is necessary for the evaluation of
$m_i(t)$. Following earlier studies
\cite{OpperWinther-2001,Moudi-2013}, we approximately replace these
with those of the full distribution as
\begin{subequations}\label{Var}
\begin{align}
V_{\eta_i \eta_i}&\simeq \left \langle \left (\sum_{l \ne i} J_{il}
(s_l(t-1) -\left \langle s_l(t-1) \right \rangle ) \right )^2 \right
\rangle \cr &\simeq \sum_{l \ne i}J_{il}^2 \left (1-\left \langle
s_l(t-1)\right \rangle^2 \right )
\simeq \frac{J^2 }{N}\sum_{l=1}^N (1-m_l^2(t-1))\equiv V_{\eta\eta}(t), \label{Vara}\\
V_{\eta_{i}\phi_{i}} & \simeq \left \langle \left (\sum_{j \ne i}
J_{ij} (s_j(t-1)-\left \langle s_j(t-1) \right \rangle ) \right )
\left (\sum_{l \ne i} J_{li} (s_l(t-1)- \tanh(\beta h_l^{(i)}(t-1) )
) \right ) \right \rangle \cr &\simeq \sum_{l\neq i} J_{il}J_{li}
\left (1 -\left\langle\tanh^{2}(\beta h_l(t-1) )\right\rangle \right
) \simeq \left (\frac{1-k^2}{1+k^2} \right )\frac{J^2}{N}
\sum_{l=1}^N
 (1\!-\!\hat{m}_l(t\!-\!1) )\nonumber\\
 & \equiv V_{\eta\phi}(t), \label{Varb}
 \end{align}
\end{subequations}
where $\hat{m}_{l}(t-1)\equiv\left \langle\tanh^{2}(\beta h_l(t-1) )
\right \rangle $ is evaluated using the update rule
\begin{equation}\label{Var-mL}
    \hat{m}_{l}(t)=\sum_{s_{l}(t-2)}\frac{1+m_{l}(t-2)s_{l}(t-2)}{2}\int
    Dz\tanh^{2}(\beta\Xi(z,s_{l}(t-2)))
\end{equation}
for the $t-1$-th step.

Equations (\ref{IMF-mag})--(\ref{Var-mL}) constitute our improved
mean field theory. In practice, this is carried out as follows:
\begin{itemize}
\item Expectations for $t=0$ and $1$ are evaluated exceptionally as
$m_i(0)=\sum_{s_i(0)} s_i(0) P_i(s_i(0))$, $m_i(1)=\int Dz \tanh
\left (\beta \Xi'_i(z,1) \right )$ and $\hat{m}_i(1)=\int Dz \tanh^2
\left (\beta \Xi'_i(z,1) \right )$ for $i=1,2,\ldots,N$, where
$\Xi'_i(z,t)\equiv\theta_{i}(t)+\sum_{j\in\partial
i}J_{ij}m_{j}(t-1)+\sqrt{V_{\eta\eta}(t)}z$. These provide the
initial condition for the subsequent dynamics.
\item For $t \ge 2$, $V_{\eta\eta}(t)$ and $V_{\eta\phi}(t)$ are computed first from $\{m_i(t-1)\}$ and $\{\hat{m}_i(t-1)\}$
by Eqs. (\ref{Vara}) and (\ref{Varb}), respectively. Then, $m_i(t)$
and $\hat{m}_i(t)$ are assessed from $m_i(t-1)$ and $m_i(t-2)$ with
the use of $V_{\eta\eta}(t)$ and $V_{\eta\phi}(t)$ following Eqs.
(\ref{IMF-mag}) and (\ref{Var-mL}).
\end{itemize}
In the above treatment, we dropped all terms negligible for $N\to
\infty$. Keeping the site dependence in Eqs. (\ref{Vara}) and
(\ref{Varb}) and/or considering the contributions from the
off-diagonal correlations as developed in the previous section may
improve the approximation accuracy for relatively small systems.
Note that, by applying the above procedure starting from
$\mathbf{m}(0)$, we can only capture the short-time trend of
dynamics (measured by the evolution of the global magnetization
(data not shown)). This suggests that we should improve the
approximation by considering correlations at more time steps.
However, under the assumption of stationarity,
$V_{\eta\phi}(t)=V_{\eta\phi}(t-1)$, $V_{\eta\phi}(t)$ can be
determined self-consistently, which is effective in practical
prediction, as we shall show subsequently.

\begin{figure}
\centering
    \includegraphics[bb=20 20 291 217,width=0.8\textwidth]{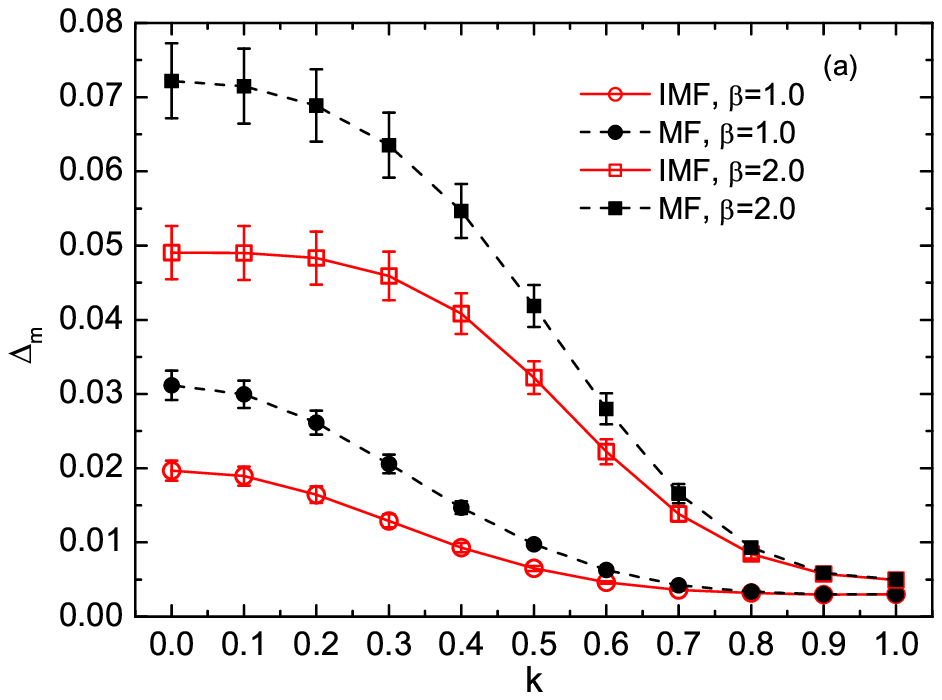}\vskip .1cm
    \includegraphics[bb=18 14 286 223,width=0.8\textwidth]{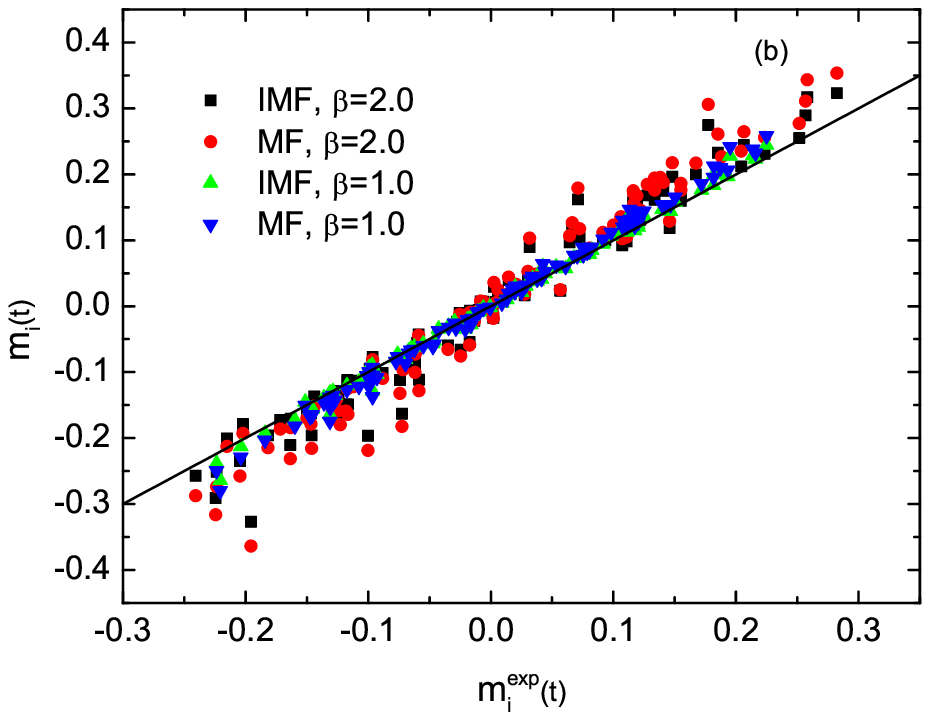}\vskip .1cm
  \caption{
  (Color online) Comparison of prediction performance of IMF and MF on partially asymmetric kinetic Ising systems of system size $N=100$.
  Each data point is the average over ten random realizations.
A total number of $10^{5}$ spin trajectories are collected up to
$31$ time steps and these trajectory data are used either to compute
inputs for prediction equations or to compute the experimental values
for comparison. We predict the instantaneous magnetization at $t=31$
based on the data at $t=30$ and $t=29$. (a) Constant external fields
with $\theta_{0}(t)=0.1$. The asymmetry degree $k$ is varied and the
result for two different temperatures is shown. (b) Scatter plot for
a typical example with $k=0.3$ in (a). The full line indicates
equality.
  }\label{IMFcomp}
\end{figure}

\begin{figure}
\centering
    \includegraphics[bb=17 13 291 219,width=0.8\textwidth]{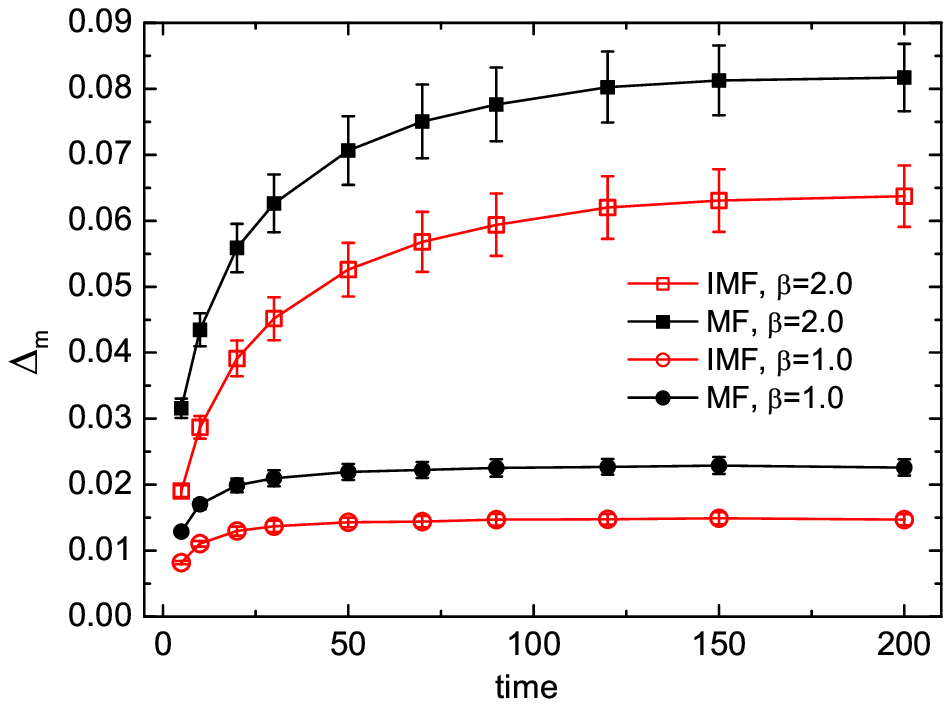}
  \caption{
  (Color online) Time dependence of the prediction performance of IMF and MF on partially asymmetric kinetic Ising systems of system size $N=100$
  and asymmetry degree $k=0.3$.
  Each data point is the average over ten random realizations with constant external fields $\theta_{0}(t)=0.1$.
  }\label{IMFtime}
\end{figure}

We remark here that $V_{\eta\phi}$ (Eq.~(\ref{Varb})) vanishes in
the fully asymmetric network and Eq.~(\ref{IMF-mag}) gives back
Eq.~(\ref{Mqa}), which is exact when the network is fully
asymmetric. However, even if the asymmetry degree $k\neq1$, our
theory is expected to have a good prediction performance as the
contributions from the backaction field $\phi_{i}$ are explicitly
considered. To examine this point clearly, we compared the
prediction error of instantaneous magnetization by using
Eqs.~(\ref{IMF-mag}) and (\ref{Mqa}) based on the numerically
collected data, which is shown in Fig.~\ref{IMFcomp}. To keep the
same low complexity as in Ref.~\cite{Mezard-2011,Mezard-2012}, we
adopt Eq.~(\ref{Vara}) by assuming the nondiagonal correlations to
be negligible. In the prediction, we kept the site dependence in
Eqs. (\ref{Vara}) and (\ref{Varb}), and we determined
$V_{\eta_i\phi_i}(t)$ on the basis of the data of $m_i(t-1)$ and
$m_i(t-2)$ in a self-consistent manner~\cite{Suppl} assuming that
the dynamics reaches the stationary state, so that
$V_{\eta_i\phi_i}(t)=V_{\eta_i\phi_i}(t-1)$ holds. As seen in
Fig.~\ref{IMFcomp}(a), IMF definitely outperforms MF, especially for
$k$ close to zero with strong coupling correlations. The improvement
becomes more apparent in the low-temperature region. As
$k\rightarrow 1$, the prediction error of both methods becomes
indistinguishable, as expected from the above theoretical
derivation. From the scatter plot in Fig.~\ref{IMFcomp}(b), one can
conclude that IMF predicts a value of magnetization closer to the
true value, compared to MF. Figure~\ref{IMFtime} explores the time
dependence of the prediction performance, which shows that IMF
always yields a better performance than MF, and the prediction error
saturates at large time for both methods. Figure~\ref{IMFtime} also
implies that, even at short time, IMF still well predicts the
experimental results.

\section{Summary}
\label{sec_Sum}

In this paper, we proposed two schemes for improving the existing
mean-field description of the dynamics of a kinetic Ising spin
model. In the first scheme, we showed that the formula for the
time-delayed correlation can be recovered without the
small-correlation (of local fields) assumption. In addition, we
developed formulas for improving the prediction accuracy of
magnetizations, the same- and delayed time correlations by
incorporating the pairwise correlations of local fields, which are
particularly effective in the low temperature region.

In the second scheme, we focused on considering the influence of
statistical correlations between couplings of two opposite
directions for each pair of spins. When statistical correlations
exist for the coupling pairs, the central limit theorem assumed in
the existing mean-field theory, which was developed by supposing a
fully asymmetric network, does not hold. Local fields of different
spins correlate with one another in a complex way, and furthermore,
the instantaneous value of spin is not independent of the couplings.
To properly treat this significant memory effect present in a
general system, we developed an improved mean-field theory utilizing
the notion of the cavity system in conjunction with a perturbative
expansion approach. Its efficiency was numerically confirmed by
comparison with the existing mean-field theory.

Note that the first scheme applies a similar idea to the recent work
by Mahmoudi and Saad~\cite{Moudi-2013}, but in their
work, the (auto-) field covariances are calculated recursively,
which may demand expensive computational cost, like the case of
MFcorre whose computational cost is of the order $N^{3}$. However,
in the second scheme, by introducing additional backaction fields
(on top of the original local fields), IMF provides efficient
predictions with low complexity ($\sim\mathcal {O}(N^{2})$, the same
as that of MF), while the usual Monte Carlo simulation takes a
computer time proportional to $N^{2}tP_{t}$ where $t$ denotes the
length of one trajectory and $P_{t}$ is the total number of
trajectories. $P_{t}$ usually takes a large value (e.g., $10^{5}$)
to ensure numerical accuracy.

 Studies of such
nonequilibrium behavior of asymmetric kinetic Ising systems could
provide insights into nonequilibrium network reconstruction, which
has received considerable interest in recent
years~\cite{Roudi-2011prl,Mezard-2011}, for example, for improving
the coupling and field inference in the context of dynamical
inference. The two schemes proposed in this paper should prove
promising for developing an inverse mean-field algorithm to
construct asymmetric couplings between elements in a network based
on time-series data.

\section*{Acknowledgments}
 This work was partially supported by the JSPS Fellowship for Foreign
Researchers (Grant No. $24\cdot02049$) (HH) and JSPS/MEXT KAKENHI
Grant Nos. $22300003$, $22300098$, and $25120013$ (YK).


\begin{thebibliography}{10}
\bibitem{Suppl}
One method to get covariance $V_{\eta_{i}\phi_{i}}(t)$ is to
iteratively update
  $\{\hat{m}_{l}(t-1)\}_{l=1}^{l=N}$ until they converge within some prescribed
  numerical precision. Note that when $\hat{m}_{l}$ is updated, all
  $V_{\eta_{i}\phi_{i}}$ $(i\neq l)$ should also be updated as
  $V_{\eta_{i}\phi_{i}}^{t+1}=V_{\eta_{i}\phi_{i}}^{t}-J_{il}J_{li}\Delta\hat{%
m}_{l}$, where $t$ denotes the iteration step and
$\Delta\hat{m}_{l}$ denotes
  the change of $\hat{m}_{l}$.
\end{thebibliography}


\end{document}